\documentclass[a4paper,12pt]{article}
\pdfoutput=1
\usepackage{feynmp-auto,expdlist}
\usepackage{amsmath, amsfonts, amssymb}
\usepackage{graphicx}
\usepackage{enumerate,cancel}
\usepackage{hyperref}
\usepackage{latexsym}
\usepackage{dsfont}
\usepackage{hepnicenames}
\usepackage{enumerate}
\usepackage{soul}
\usepackage[normalem]{ulem}
\usepackage{comment}
\usepackage{units}

\newcommand{\via}[1]{\footnote{\tiny #1}}
\renewcommand{\via}[1]{{}}
\usepackage{soul}

\newcommand{\bAk}[3]{\langle #1|#2|#3\rangle}

\usepackage{mathrsfs,graphicx,rotating,amsmath,amsfonts,mathtools,booktabs,amssymb,wasysym}
\usepackage{hyperref}\usepackage{slashed}
\usepackage[nosort]{cite}
\usepackage[table,xcdraw]{xcolor}
\usepackage{bm}
\usepackage{graphicx}
\usepackage{multirow,multicol}
\hypersetup{colorlinks,bookmarksopen,bookmarksnumbered,
	linkcolor=blus,pdfstartview=FitH,urlcolor=rossos,citecolor=verde}
\allowdisplaybreaks

\newcommand{\mDG}{M_{\scriptscriptstyle{\rm DG}}}

\renewcommand{\[}{\left[}

\newcommand{\Lag}{\mathscr{L}}

\newcommand{\mio}[1]{}

\newcommand{\med}[1]{\langle #1\rangle}

\newcommand{\bpm}{\begin{pmatrix}}
\newcommand{\epm}{\end{pmatrix}}

\usepackage{mathrsfs}

\newcommand{\fig}[1]{~\ref{fig:#1}}

\allowdisplaybreaks
\usepackage{multicol}
\usepackage{color}
\definecolor{rosso}{cmyk}{0,1,1,0.4}
\definecolor{rossos}{cmyk}{0,1,1,0.55}
\definecolor{rossoc}{cmyk}{0,1,1,0.2}
\definecolor{blu}{cmyk}{1,1,0,0.3}
\definecolor{blus}{cmyk}{1,1,0,0.6}
\definecolor{bluc}{cmyk}{1,1,0,0.1}
\definecolor{verde}{cmyk}{0.92,0,0.59,0.25}
\definecolor{verdec}{cmyk}{0.92,0,0.59,0.15}
\definecolor{verdes}{cmyk}{0.92,0,0.59,0.4}

\newcommand{\bp}{\bar{M}_{\rm Pl}}
\newcommand{\eq}[1]{~{\rm (\ref{eq:#1})}}

\newcommand{\MeV}{\,{\rm MeV}}
\newcommand{\GeV}{\,{\rm GeV}}

\newcommand{\Tr}{\,{\rm Tr}}
\newcommand{\diag}{\,{\rm diag}}

\newcommand{\beq}{\begin{equation}}
\newcommand{\eeq}{\end{equation}}

\newcommand{\bea}{\begin{eqnarray}}
\newcommand{\eea}{\end{eqnarray}}
\newcommand{\be}{\begin{equation}}
\newcommand{\ee}{\end{equation}}
\font\tenrsfs=rsfs10 at 12pt
\font\sevenrsfs=rsfs7
\font\fiversfs=rsfs5
\newfam\rsfsfam
\textfont\rsfsfam=\tenrsfs
\scriptfont\rsfsfam=\sevenrsfs
\scriptscriptfont\rsfsfam=\fiversfs
\newcommand{\mpl}{M_{\rm Pl}}
\newcommand{\tdec}{T_{\rm decay}}
\newcommand{\trh}{T_{\rm RH}}
\newcommand{\rdg}{\rho_S}

\def\be#1\ee{\begin{equation}#1\end{equation}}
\def\bl#1\el{\begin{align}#1\end{align}}
\def\ba#1\ea{\begin{align*}#1\end{align*}}

\renewenvironment{thebibliography}[1]
{\begin{multicols}{2}[\section*{\refname}]%
		\@mkboth{\MakeUppercase\refname}{\MakeUppercase\refname}%
		\list{\@biblabel{\@arabic\c@enumiv}}%
		{\settowidth\labelwidth{\@biblabel{#1}}%
			\leftmargin\labelwidth
			\advance\leftmargin\labelsep
			\@openbib@code
			\usecounter{enumiv}%
			\let\p@enumiv\@empty
			\renewcommand\theenumiv{\@arabic\c@enumiv}}%
		\sloppy
		\clubpenalty4000
		\@clubpenalty \clubpenalty
		\widowpenalty4000%
		\sfcode`\.\@m}
	{\def\@noitemerr
		{\@latex@warning{Empty `thebibliography' environment}}%
		\endlist\end{multicols}}

\newcommand{\eV}{\,{\rm eV}}
\newcommand{\SU}{\,{\rm SU}}
\newcommand{\SO}{\,{\rm SO}}

\makeatletter
\font\ital=cmu10

\def\hhref#1{\href{http://arxiv.org/abs/#1}{arXiv:#1}}
\usepackage{xstring}
\newcommand{\hhrefq}[1]{\IfSubStr{#1}{:}{\href{http://inspirehep.net/search?ln=en&ln=en&p=#1&of=hb&action_search=Search&sf=&so=d&rm=&rg=25&sc=0}{InSpire:#1}}{\hhref{#1}}}

\def\art{\@ifnextchar[{\eart}{\oart}}
\def\eart[#1]#2#3#4#5#6{{\rm #2}, {\em #3 \bf #4} {\rm (#6) #5} ({\em #1})}
\def\article{\@ifnextchar[{\earticle}{\oarticle}}
\def\oarticle#1#2#3#4#5#6{{\rm #1}, {\ital `#6'}, {\rm #2 #3 (#5) #4}}
\def\earticle[#1]#2#3#4#5#6#7{{\rm #2}, {\ital `#7'}, {\rm #3 #4 (#6) #5}  [\hhrefq{#1}]}
\def\hepart[#1]#2{{\rm #2, \sl#1}}
\def\heparticle[#1]#2#3{#2, {\ital `#3'} [\hhrefq{#1}]}
\newcommand{\doi}[1]{\href{http://dx.doi.org/#1}{[link]}}

\newcommand{\hhrefqq}[1]{\IfBeginWith{#1}{10.}{\href{https://doi.org/#1}{doi:#1}}{\hhrefq{#1}}}

\renewenvironment{thebibliography}[1]
{\begin{multicols}{2}[\section*{\refname}]%
		\@mkboth{\MakeUppercase\refname}{\MakeUppercase\refname}%
		\list{\@biblabel{\@arabic\c@enumiv}}%
		{\settowidth\labelwidth{\@biblabel{#1}}%
			\leftmargin\labelwidth
			\advance\leftmargin\labelsep
			\@openbib@code
			\usecounter{enumiv}%
			\let\p@enumiv\@empty
			\renewcommand\theenumiv{\@arabic\c@enumiv}}%
		\sloppy
		\clubpenalty4000
		\@clubpenalty \clubpenalty
		\widowpenalty4000%
		\sfcode`\.\@m}
	{\renewcommand{\@noitemerr}
		{\@latex@warning{Empty `thebibliography' environment}}%
		\endlist\end{multicols}}

\newcounter{alphaequation}[equation]
\renewcommand{\thealphaequation}{\theequation\hbox to
	0.6em{\hfil\alph{alphaequation}\hfil}}
\definecolor{Gray}{gray}{0.95}

\usepackage{tocloft}
\begin{document}
\thispagestyle{empty}
\begin{center}  
{\LARGE\bf\color{rossos}Optical gravitational waves as signals\\[0.5ex] 
of Gravitationally-Decaying Particles } \\[3ex]
 {\bf Alessandro Strumia}$^a$, {\bf Giacomo Landini}$^b$   \\[2ex]
{\it $^a$ Dipartimento di Fisica, Universit\`a di Pisa, Italia}\\[1ex]
{\it $^b$ Instituto de F\'isica Corpuscular, Universitat de Val\`encia-CSIC, Spain}\\[0.5ex]
\vspace{0.5cm}
{\large\bf Abstract}
\begin{quote}
Long-lived heavy particles present during the big bang
could have a decay channel opened by gravitons.
Such decays can produce gravitational waves
with large enough abundance to be detectable,
and a peculiar narrow spectrum peaked today around optical frequencies.
We identify which particles can decay in one or two gravitons.
The maximal gravitational wave abundance arises from theories with extra hidden strong gauge dynamics, such as a 
confining pure-glue group.
An interesting abundance also arises in theories with perturbative couplings.
Future observation might shed light on early cosmology
and allow some spectroscopy of sub-Planckian 
gravitationally-decaying particles, plausibly present in a variety of theories such as
gauge unification, supersymmetry, extra dimensions, strings.
\end{quote}
\end{center}
\tableofcontents

\newpage

\section{Introduction}
\begin{figure*}[t]
$$\includegraphics[width=0.9\textwidth]{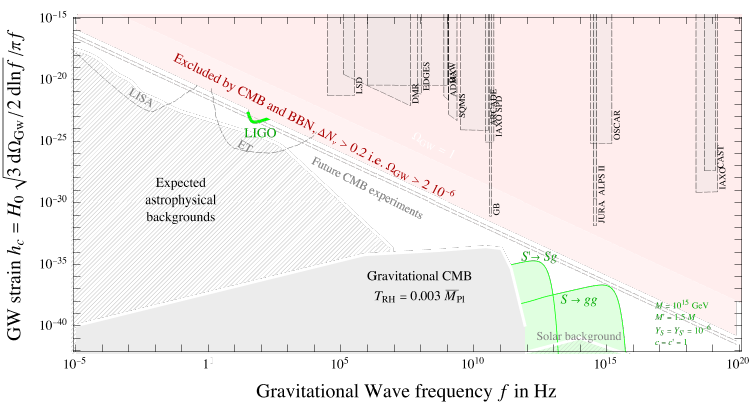}$$
\vspace{-5ex}
\caption{\em \label{fig:GW2} 
Panoramic plot.
The vertical axis is the gravitational wave characteristic strain $h_c$, connected to the energy density by
$d\Omega_{\rm GW}/d\ln f=2\pi^2 f^2 h_c^2/3H_0^2$.
The red region is excluded by the CMB/BBN bound of eq.\eq{Neff_bound}.
The continuous curve is the LIGO/VIRGO GW observation.
The dashed curves are sensitivities of possible future experiments with the indicated names.
The gray region is the maximal cosmic microwave gravitational background.
The hatching indicates expected astrophysical backgrounds, including the solar background at optical frequencies~\cite{gr-qc/0406089,2407.18297}.
The two green peaks at optical frequencies show an example of
optical gravitational waves from decays of the spin 0 and spin 2 glue-balls $S$ and $S'$
 in the theory of section~\ref{sec:G}.
} 
\end{figure*}

Gravitons interact much less than neutrinos or other SM particles, 
so relic cosmological gravitational waves would provide information on the very early universe.
However, the tiny interaction renders challenging detecting gravitational waves.
The average cosmological energy density $d\rho_{\rm GW}/d\ln f$
of relic gravitons $g$ with frequency $f$ must satisfy CMB and BBN constraints that exclude 
an extra relativistic species with abundance comparable to neutrinos~\cite{Planck:2018vyg}:
\beq
\Delta N_{\nu}^{\rm eff}\equiv 1.8~10^{5}\int  h^2 \frac{d\Omega_{\rm GW}}{d\ln f}d\ln f \lesssim  0.2\qquad
\hbox{where}\qquad\frac{d\Omega_{\rm GW}}{d\ln f} \equiv  \frac{1}{\rho_{\rm cr}} \frac{d \rho_{\rm GW}}{d\ln f},
\label{eq:Neff_bound}
\eeq
$\rho_{\rm cr}=3 \bp^2 H_0^2$, 
$H_0= h ~ 100\,{\rm km/s}\,{\rm Mpc}$,
$h \approx 0.67$. Future CMB data are expected to improve the sensitivity by one order of magnitude
thanks to higher precision.
A cosmological detection of some extra radiation would not tell its identity nor its spectrum.
More interestingly, future experiments based on graviton to photon conversion in magnetic fields
aim at detecting gravitational waves with good spectral resolution 
around atomic frequencies $f\sim 10^{15-19}\,{\rm Hz}$.
For the moment, experiments planned to search for axions using magnets built for colliders
use a static magnetic field $B$ in an area $S$ and length $L \lesssim S/\lambda_{\rm GW}$ to maintain coherence:
gravitational waves induce a photon power $W_\gamma \sim \rho_{\rm GW} S L^2 B^2/\bp^2$.
This is below the minimal detectable power if $\Omega_{\rm GW}\lesssim 1$~\cite{Ballantini:2005am,2011.12414,Arvanitaki:2012cn,Ejlli:2019bqj,Aggarwal:2020umq,LSD:2022mpz,Berlin:2021txa,Goryachev:2021zzn,Campbell:2023qbf,Sorge:2023nax,Tobar:2023ksi,Domcke:2022rgu,Bringmann:2023gba,Vacalis:2023gdz,Liu:2023mll,Ito:2019wcb,Gatti,2011.04731,Ito:2022rxn,Ito:2023bnu,2203.15668,2304.11222,2311.17147,2412.14450,2501.11723}, as illustrated in fig.\fig{GW2}.\footnote{These experiments are thereby
not sensitive to cosmological sources, only to gravitational waves favourably localised in time or space.
As an aside comment we mention the following possibility: dark matter has a galactic density $10^5$ times higher than
the cosmological density and might decay in gravitons only with life-time $\tau\gtrsim 250\,{\rm Gyr}$~\cite{2203.07440},
thereby generating a local $\Omega_{\rm GW}\sim 10^5 R/\tau \lesssim 0.1$ where
$R\sim 10^5\,{\rm yr}$ is the galactic size.}
Opening a new observational window requires reaching sensitivities below the BBN/CMB bound of eq.\eq{Neff_bound}.
Innovation will be needed: adding a laser beam with electric field $E$ oscillating
at the same frequency as the gravitational waves
induces a signal linear in the gravitational wave amplitude, $W_\gamma \sim \sqrt{\rho_{\rm GW}} SL BE/\bp$, 
but also introduces a background~\cite{gr-qc/0306092,2011.04731}.

%


\medskip

Various processes could produce a cosmological background of
relic gravitons with $f \sim T_0 \sim 300\,{\rm GHz}$, the current CMB temperature.
However, most processes produce a graviton abundance significantly lower than that of photons or neutrinos. 
This scarcity stems from the suppression of graviton couplings
$E/\bp$ at energies $E$ below the reduced Planck mass $\bp\approx 2.4~10^{18}\GeV$.  
The minimal contribution, from scatterings among SM particles, produces a gravitational cosmological background 
with abundance
$\Delta N_{\nu}^{\rm eff} \approx 
0.01 T_{\rm RH}/\bp \lesssim 10^{-4.5}$ around microwave frequencies~\cite{2004.11392},
as illustrated by the gray region in fig.\fig{GW2}.
This is small because the reheating temperature must be below $T_{\rm RH} \lesssim 0.003 \bp$
to avoid overproducing inflationary tensor modes~\cite{Planck:2018vyg}.

\smallskip

This paper discusses an exception to the typical outcome of a small graviton relic background:
some particle $S$ with mass $M \sim 10^{10-18}\GeV$ 
might have been present during the big-bang and decayed slowly.
Gravitons might have {\em opened} a channel for its decay
(as opposed to having a fast decay channel 
already opened by other particles and adding a small
branching ratio $\sim (M/4\pi \bp)^2$ into gravitons with
smooth spectrum peaked around micro-waves~\cite{1810.04975,2211.10433,2310.12023,2311.12694,2312.16691,2403.13882,2407.03256,2410.01900}).
In view of the assumed slow decay, 
small Planck-suppressed graviton couplings can lead to a big branching ratio into gravitons,
as they only compete with other small couplings.
Moreover, during the big bang, a weakly interacting 
long-lived massive particle acquires an enhanced abundance,
because matter gets diluted more slowly than SM radiation.
As a result, the resulting graviton abundance $\Omega_{\rm GW}$ can easily approach the bound in eq.\eq{Neff_bound}.

In section~\ref{sec:cosmo} we compute the assumed cosmology and the resulting
peculiar spectrum of produced gravitons, peaked at $f\gg T_0$ in view of the slow decay. 
In section~\ref{sec:Amp} we apply helicity methods to classify which processes 
$S\to gg$, $S\to g\,{\rm SM}$, $S'\to S g$ are allowed by Lorentz invariance and locality,
where $S,S'$ are massive particles, and SM denotes any SM particle.
In section~\ref{sec:models} we discuss specific theories.
An hidden sector with strong coupling provides the maximal graviton abundance.
Theories with perturbative couplings (in particular, extensions of Einstein gravity)
can plausibly contain one or more particles that slowly decay
gravitationally producing detectable gravitational waves.
Observing a multi-peaked spectrum of optical gravitational waves could
allow a spectroscopy of such states, offering an experimental window into
theories such as gauge unification, supersymmetry, extra dimensions, string theories, and more.
Conclusions are given in section~\ref{concl}.

\section{A particle decaying into gravitons}\label{sec:cosmo}
We here compute the cosmology during the big-bang  at temperature $T$
of a particle $S$ with mass $M$, 
number abundance $n_S $ 
and slow decay rate 
$\Gamma_S= \Gamma_{\rm GW} + \Gamma_{\rm SM} \ll M$ 
with a significant $ \hbox{BR}_{\rm GW}= \Gamma_{\rm GW}/\Gamma_S $ {\em opened} by one or more gravitons
(as opposed to adding a graviton to an allowed decay into SM particles).
We parameterise the $S$ decay width into gravitons as
\beq\label{eq:GammaGW}
 \Gamma_{\rm GW}= \frac{cM}{4\pi} \left(\frac{M}{\bp}\right)^p   \eeq 
where $c$ and $p$ are dimension-less constants and $\bar{M}_{\rm Pl}=M_{\rm Pl}/\sqrt{8\pi}=2.4~10^{18}\GeV$
is the reduced Planck mass.
As discussed in section~\ref{sec:models}, the plausible values that maximise $\Gamma_{\rm GW}$ seem $p=2$
(for one graviton production) or $p=4$ (for two-graviton production) with
$c\sim 1$. 
This maximal operator coefficient arises out of minimal graviton couplings in strongly-coupled theories with one scale.
In general, $c$ can be suppressed by a one loop factor (which is small in theories with perturbative couplings),
as well as by the mass of particles in the loop, in case they are heavier than $S$ itself.

The minimal gravitational decay rate into SM particles 
due to virtual one-graviton exchange or related gravitational operators
is expect to be suppressed
by 4 powers of the Planck mass,
\beq \label{eq:GammaSM}
\Gamma_{\rm SM}  \gtrsim \frac{g_{\rm SM} M}{4\pi}  \left(\frac{M}{\bp}\right)^4 \eeq
where $g_{\rm SM}\approx 106.75$ is the number of SM degrees of freedom.
Extra particles could exist at energies around $M$.
Of course, extra non-gravitational interactions can enhance $\Gamma_{\rm SM}$.

\subsection{Production of $S$}
We here consider plausible values of the cosmological number density $n_S$ of $S$.
It is convenient to consider the combination $Y_S = n_S/s$ not affected by the expansion 
thanks to the entropy density $s = 2\pi^2 g_{\rm SM}T^3/45$.
Thermal equilibrium corresponds to $Y_S\sim 1/g_{\rm SM}$.
A minimal contribution to $Y_S$ arises from
the inverse process ${\rm SM}\,{\rm SM}\to S$ with space-time rate density $\gamma \sim s \Gamma_{\rm SM}e^{-M/T}$.
This inverse decay contributes to the $S$ abundance as:
\beq \label{eq:YSmin}
Y_S \equiv \frac{n_S}{s}\sim \max_T  \frac{\gamma}{Hs} \sim \Gamma_{\rm SM} \frac{\bp}{M^2}e^{-M/T_{\rm RH}} \sim \left(\frac{M}{\bp}\right)^3
e^{-M/T_{\rm RH}} \eeq
having assumed eq.\eq{GammaSM}.
The peculiar inverse decay kinematics does not allow to benefit from the possibility that
the reheating temperature $\trh$ after inflation is larger than $M$.
So a larger contribution to the $S$ number density 
can arise from gravitational ${\rm SM}\,{\rm SM}\to SS$ pair production, 
with rate $\gamma\sim T^8/\bp^4$ resulting in
\beq \label{eq:YSmin2}
Y_S\sim 10^{-2-3}\left(\frac{\trh}{\bp}\right)^3e^{-M/\trh}.\eeq
This contribution can be larger in specific theories with extra interactions.
Furthermore, inflationary production (considered in~\cite{2112.12774}) can contribute to $Y_S$,
and $S$ particles heavier than $T_{\rm RH}$ can be produced if their mass depends on the
inflaton vacuum expectation value and get accidentally lighter during a temporary phase near to the end of inflaton.
We will keep $Y_S$ as a free parameter.

\subsection{Decay of $S$}
We here compute the total gravitational wave abundance $\rho_{\rm GW}$
produced from $S$ decays.
Since the decay is slow, $S$ decays while non-relativistic and out of equilibrium.
We assume that its interactions are small enough to be cosmologically negligible.
The Hubble rate at temperature $T$ and time $t$ is
\begin{equation}\label{eq:Hubble}
H^2=\frac{\rho_{\rm SM}+\rho_{S}}{3 \bp^2},\qquad   \rho_{\rm SM} = \frac{\pi^2}{30}g_{\rm SM}(T)T^4.
\end{equation}
The cosmological evolution during $S$ decays is described by
\beq \label{eq:cosmoevo}
\frac{d\rho_S}{d\ln a}=-3\rho_S - \frac{\Gamma_S}{H}\rho_S,\qquad
\frac{d\rho_{\rm SM}}{d\ln a}= -4\rho_{\rm SM} +\frac{\Gamma_{\rm SM}}{H}\rho_S,\qquad
\frac{d\rho_{\rm GW}}{d\ln a}= -4\rho_{\rm GW} +\frac{\Gamma_{\rm GW}}{H}\rho_S.
\eeq
The equations for $\rho_S$ and  $\rho_{\rm GW}$
are solved starting at an initial $a=a_i$ by 
\beq\label{eq:rhoanal}
\rho_{S}(a)=\rho_S(a_i)  \bigg(\frac{a_i}{a}\bigg)^3   \, e^{-\Gamma_S t},\qquad
\rho_{\rm GW}(a) =\int_{a_i}^a d\ln a'  \bigg(\frac{a'}{a}\bigg)^4 \frac{\Gamma_{\rm GW}\rho_S}{H}.\eeq
Eq.s\eq{cosmoevo} can be rewritten omitting the 2nd equation and 
switching as variable from scale factor $a$ to temperature $T$
(defined in terms of $\rho_{\rm SM}$ as in eq.\eq{Hubble}),
by using (see e.g.~\cite{hep-ph/0310123}):
\beq  Z= - \frac{d\ln T}{d\ln a}
= 1 - \frac14 \frac{\Gamma_{\rm SM}\rho_S}{H\rho_{\rm SM}} .
\eeq
So $T \propto 1/a^Z$. Initially $Z\simeq 1$ even when $\rho_S >\rho_{\rm SM}$.
Next, after $\rho_S/\rho_{\rm SM} > H/\Gamma_{\rm SM}$, 
one has  $Z\simeq 3/8$ i.e.\ $T \propto a^{-3/8}$ cools more slowly,
due to reheating from $S$ decays.
Reheating ends when $1/t\sim H \sim \Gamma_S$, reverting to $Z\simeq 1$.

\begin{figure*}[t]
$$\includegraphics[width=0.7\textwidth]{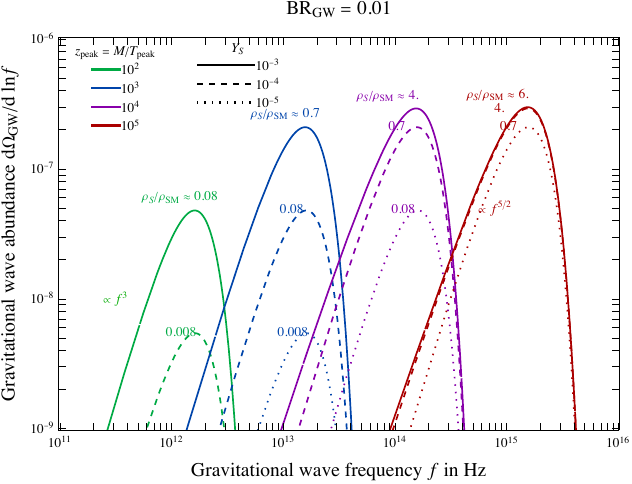}$$
\caption{\em \label{fig:GWdecaySpectra} 
Spectra of gravitational waves produced by a scalar $S$ 
with mass $M$ and initial abundance $Y_S$ 
slowly decaying into two gravitons at the temperature $T_{\rm peak}=M/z_{\rm peak}$
with branching ration into gravitons $\hbox{\rm BR}_{\rm GW}=0.01$.
The plot illustrates the full parameter space: $z_{\rm peak}$ accounts for both the scalar mass and decay width,
and different values of the branching ratio in gravitons simply increase or decrease the overall gravitational wave abundance.
The gravitational wave signals lie below the CMB/BBN bound  of eq.~\eqref{eq:Neff_bound} in all the plotted region.
The numbers indicate $\rho_S/\rho_{\rm SM}$ at peak for each spectrum.} 
\end{figure*} 

\subsection{Gravitational wave spectrum}
While SM particles (including neutrinos at $T \gtrsim \MeV$) interacted and thermalised,
gravitons would have travelled nearly free, such that
the relic gravitational wave spectrum today still reflects early cosmology.
We write as $dN/d\ln E$ the differential spectrum of gravitons with energy $E$ produced by each $S$ decay.
Assuming that it happens at scale factor $a$, 
gravitons have current energy $E_0=aE$.
So the current graviton energy density spectrum is
\beq \frac{d\rho_{\rm GW}}{d\ln E_0}= E_0 \int \frac{da}{a} a^3 \frac{\rho_S}{M}  \frac{\Gamma_{\rm GW}}{H} \frac{dN}{d\ln E}.
\eeq
For example, $S\to gg$ decays correspond to $dN/dE = 2\delta(E-M/2)$, such that
the graviton spectrum is a line convoluted with the exponential distribution of decay times, that gives different redshift:
\beq \frac{d\rho_{\rm GW}}{d\ln E_0}= a^4 \rho_S \frac{\Gamma_{\rm GW}}{H}  \qquad\hbox{evaluated at}\qquad
 a = \frac{2E_0}{M}.\eeq 
In view of the simplest particle physics (a line),
the frequency $f = E_0/2\pi$
spectrum of gravitational waves directly reflects different epochs of the cosmology during $S$ decays.
In our case $d\Omega_{\rm GW}/d\ln f$ has a peak with order unity width at some frequency $f_{0\rm peak}$.

The gravitational wave spectra can be computed in terms of three effective parameters: 
\begin{itemize}
\item[1)] the $S$ abundance $Y_S$;
\item[2)] the $S$ branching ratio into gravitons $\hbox{BR}_{\rm GW}$, that acts as a proportionality constant;
\item[3)] the dimension-less ratio $z_{\rm peak}\equiv{M}/{T_{\rm peak}} \gg 1$ between the $S$ mass and the 
temperature  $T_{\rm peak} $ that solves $\Gamma_S \approx H(T_{\rm peak})$,
at which the $S$ decay dominantly happens.
\end{itemize}
At this temperature, $S$ can either be
a negligible or dominant component of the cosmological energy density,
depending on whether the $S$ density during the peak of $S$ decays
is dominant or sub-dominant compared to the energy density of the SM bath:
\beq\left.\frac{\rho_S}{\rho_{\rm SM}}\right|_{\rm peak}  \sim Y_S z_{\rm peak}.  \eeq
Numerical spectra are shown in fig.\fig{GWdecaySpectra} for $\hbox{BR}_{\rm GW}=0.01$
and for different values of $z_{\rm peak}$ and of $Y_S$.
The main features of the spectra can be understood as follows.
Above the peak at $f\gg f_{0\rm peak}$ the spectrum is proportional to
an exponential cut-off $\exp[-f^{3(1+w)/2}]$,
with $w=0$ if $S$ dominates and $w=1/3$ if SM radiation dominates.
Below the peak at $f\ll f_{0\rm peak}$ the GW spectrum gets suppressed as $a^4\rho_S/H \propto a/H$.
To evaluate this factor and compute $f_{0\rm peak}$ we need to consider the two limiting cases.
\begin{itemize}
\item If $S$ decays while sub-dominant, the low-frequency tail
of the gravitational wave spectrum is suppressed by $a/H \propto a^3\propto f^3$.
The total $\rho_{\rm GW}$ will be computed in
instantaneous decay approximation in section~\ref{sec:sub}.
\item
If instead $S$ decays as dominant component and
reheats the SM plasma,
the lower-frequency tail is initially suppressed as $a/H\propto a^{5/2}\propto f^{5/2}$, implying a mildly broader spectrum around the peak.
The total $\rho_{\rm GW}$ is approximated in section~\ref{sec:dom}.
\end{itemize}
The half peak is at $f = \{0.565,1.51\}f_{0\rm peak}$ if $S$ decays while sub-dominant,
and at $f = \{0.519,1.56\}f_{0\rm peak}$ if $S$ decays while dominant.
If a positive signal will be seen, the peak frequency will tell $z_{\rm peak}=M/T_{\rm peak}$,
but data alone will not allow to separately infer $M$ and $T_{\rm peak}$.


\subsection{Decay of $S$ as sub-dominant component}\label{sec:sub}
Assuming that $S$ decays while $\rho_{\rm SM} \gg\rho_S$ dominates the energy budget, 
the decay temperature $T_{\rm peak}$ is determined by
$\Gamma_S \approx H \approx \sqrt{\rho_{\rm SM}/3}/\bp$, giving
\beq \label{eq:R}
R \equiv \frac{\rho_S}{\rho_{\rm SM}} \approx 0.77 g_{\rm SM}^{1/4} \frac{Y_S M}{\sqrt{\bp \Gamma_S}}\ll 1
\qquad
\hbox{at}\qquad T_{\rm peak}\approx M  \frac{4 Y_S}{3R}.\eeq
The graviton abundance at peak is
\beq  \label{eq:rhopeaksub}
\left.\frac{\rho_{\rm GW}}{\rho_{\rm SM}}\right|_{\rm peak}
\approx
\frac{\hbox{BR}_{\rm GW}\rho_S}{\rho_{\rm SM}} \approx 
\hbox{BR}_{\rm GW} R = \hbox{BR}_{\rm GW} \frac{4Y_S}{3} z_{\rm peak}.
\eeq
The exact solution of eq.\eq{rhoanal} can be computed analytically finding
that $\rho_{\rm GW}/\rho_{\rm SM}$ is 1.25 higher than in the instantaneous decay approximation of eq.\eq{rhopeaksub}.
Gravitons produced with energy $E=M/2$ and thereby frequency
$f = E/2\pi $ 
have today (at $T= T_0 \approx 300\,{\rm GHz}$)
a frequency redshifted down to
\beq f_{0\rm peak} \simeq \frac{M}{4\pi}  \frac{T_0}{T_{\rm peak}} = 
 \frac{T_0}{4\pi} z_{\rm peak} .\eeq  
The gravitational decays of eq.\eq{GammaGW}
predict $z_{\rm peak}\approx 2 g_{\rm SM}^{1/4} (\bp/M)^{(p-1)/2}/\sqrt{c}$.
This means that a slow decay, $z_{\rm peak}\gg 1$, can enhance $\rho_{\rm GW}$ up to its maximal allowed value,
and that the same $z_{\rm peak}$ increases the graviton frequency $f_{\rm 0peak}$ above $T_0$.

\smallskip

To conclude we convert a generic $\rho_{\rm GW}/\rho_{\rm SM}$ at a generic high temperature $T_{\rm peak}$
 into $\Omega_{\rm GW}$ today.
Gravitons redshift as radiation, $\rho_{\rm GW}\propto 1/a^4$, 
while the SM bath cools keeping its entropy $sa^3$ constant with $g_{s0}=3.91$ today.
So the current graviton energy density is
\beq\label{eq:OmegaGW}
\frac{\rho_{\rm GW}(T_0)}{\rho_{\rm GW}(T)} = \left(\frac{g_{s0}}{g_{\rm SM}}\right)^{4/3} \frac{T_0^4}{T^4},\qquad
\hbox{i.e.}\qquad
\Omega_{\rm GW} = \frac{\rho_{\rm GW}(T_0)}{\rho_{\rm cr}} =
3.6~10^{-5}  \left.\frac{\rho_{\rm GW}}{\rho_{\rm SM}}\right|_{\rm peak}
\eeq
having used $T_0 = 2.725\,{\rm K}$,
$\rho_{\rm cr}=3H_0^2 \bp^2$ with $H_0 = h\, 100 {\rm km/s Mpc}$ and $h\approx 0.67$.

\subsection{Decay of $S$ as dominant component}\label{sec:dom}
Assuming that $S$ decays while dominating the energy budget, 
the decay temperature $T_{\rm decay}$ is determined by
$\Gamma_S \approx H \approx \sqrt{\rho_S/3}/\bp$.
We obtain
\beq \frac{\rho_S}{\rho_{\rm SM}} \approx R^{4/3} \gg 1
 \qquad
 \hbox{at}\qquad 
 T_{\rm decay} \approx M \frac{4Y_S}{3R^{4/3}} 
 \eeq
with the same $R$ as in eq.\eq{R}.
The value of $T_{\rm decay}$ is however irrelevant, because
$S$ decays later reheat the SM plasma up to $T_{\rm peak}$ given by
\begin{equation}
\rho_{\rm SM}(T_{\rm peak})\equiv \frac{\pi^2}{30}g_{\rm SM}(T_{\rm peak})T_{\rm peak}^4=\rho_{\rm SM}(\tdec)+ \hbox{BR}_{\rm SM}  \,\rdg(\tdec).
\end{equation}
So the reheating temperature after $S$ decays is
 \beq  \frac{T_{\rm peak}}{M} \approx   (\hbox{BR}_{\rm SM}+R^{-4/3})^{1/4} \frac{4 Y_S}{3 R}.\eeq
It  differs from what $T_{\rm peak}$ was in the previous case of section~\ref{sec:sub}
only because the extra first factor.
The single effective parameter $z_{\rm peak}=M/T_{\rm peak} > 1$ controls
the current frequency of gravitational waves produced with energy $E=M/2$ at $T_{\rm peak}$:
\beq
 f_{0\rm peak} = \frac{ M}{4\pi} \frac{T_0}{T_{\rm peak}} \approx \frac{ T_0 z_{\rm peak}}{4\pi} 
 \eeq 
In the limit of strong $S$ dominance the graviton density at $T_{\rm peak}$ is 
\beq  \left.\frac{\rho_{\rm GW}}{\rho_{\rm SM}}\right|_{\rm peak} \simeq \frac{\hbox{BR}_{\rm GW}}{\hbox{BR}_{\rm SM}}.\eeq
and the current $\Omega_{\rm GW}$ is obtained using eq.\eq{OmegaGW}.
BBN/CMB data demand $\rho_{\rm GW}/\rho_{\rm SM}<0.03$ at peak.
This is close to the naive expectation 
${\hbox{BR}_{\rm GW}}/{\hbox{BR}_{\rm SM}} \sim 0.01$
based on assuming a common gravitational
rate and counting the number of degrees of freedom is just below current BBN/CMB bounds.

\medskip

\begin{figure*}[t]
$$\hspace{-2ex}\includegraphics[width=1.05\textwidth]{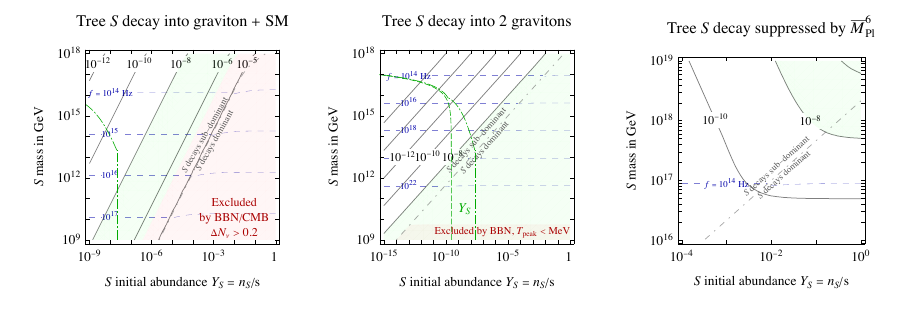}$$
\vspace{-5ex}
\caption{\em \label{fig:GW} Parameter space for the decay of a particle $S$ with mass $M$ and cosmological
abundance $Y_S=n_S/s$ into into SM particles with rate $\Gamma_{\rm SM}= g_{\rm SM}M^5/4\pi\bp^4$,
and into $N_g$ gravitons with tree-level rate $\Gamma_{\rm GW}=  M (M/\bp)^{2N_g}/4\pi$ 
(for $N_g=1$ half of the energy goes into SM particles).
$S$ decays while dominating the cosmological energy density below the grey dot-dashed line.
The continuous curves are iso-contours of the graviton abundance $\Omega_{\rm GW}$,
with the region excluded by the BBN/CMB bound of eq.\eq{Neff_bound} shaded in red.
Signals are more prominent in the region shaded in green, just below the bound.
The dashed horizontal lines are iso-contours of the current peak frequency of gravitons.
The dashed green curve is the  minimal $S$ abundance, estimated as the sum of eq.s\eq{YSmin} and\eq{YSmin2}, 
dominated by the second.
The dot-dashed green curve is the $S$ abundance in the glue-ball model of section~\ref{sec:G}.
Both are computed for the maximal $T_{\rm RH}=0.003\bp$.
The bottom corner of the middle plot is excluded by entropy injection after BBN.} 
\end{figure*}

We provided simple analytic expressions for the gravitational wave abundance and spectrum
in terms of the effective parameters $Y_S$, $z_{\rm peak}$, $\hbox{BR}_{\rm GW}$.
Finally, we assume the decay rates of eq.\eq{GammaGW} and eq.\eq{GammaSM} for different values of $p=2$
(decay into a single-graviton), $p=4$ (decay into 2 gravitons), 
$p=6$ (decay into 3 gravitons, or into 2 if the effective operator of 
eq.\eq{fund} is pessimistically suppressed by $\Lambda=\bp$).
Fig.\fig{GW} shows the resulting
iso-contours of the overall abundance $\Omega_{\rm GW}$ of gravitational waves
(computed numerically solving eq.\eq{cosmoevo}), and of the peak frequency $f_{0\rm peak}$.
The $S$-decay mechanism for GW production is so efficient that part of the parameter space (shaded in red) is already excluded.
An $\Omega_{\rm GW}$ within two orders of magnitude below current bounds arises in the area shaded in green.
Gravitationally decaying particles below the Planck scale down to about $10^{10}\GeV$ (depending on $p$)
can lead to detectable gravitational waves around optical frequencies.
For $p=\{2,4\}$ even the minimal $S$ abundance, given by the sum of eq.\eq{YSmin} and eq.\eq{YSmin2}  (dashed green curve),
allows for significant gravitational signals, provided that the reheating temperature is large enough.
The dot-dashed green curve shows the larger $Y_S$ arising in the theory of section~\ref{sec:G}.
We demand $T_{\rm peak}\gtrsim \MeV$ to avoid entropy injection after BBN. (In line of principle
the opposite extreme where $S$ is decaying today is allowed if $\hbox{BR}_{\rm SM}$ is tiny enough).

\section{Allowed decays into gravitons}\label{sec:Amp}
We here consider which $1\to 2$ decays of a massive particle allow for graviton production 
compatibly with locality and conservation of quantum numbers related to Lorentz invariance.
The Landau-Yang theorem forbids the decay of a massive vector into two identical massless vectors.
The extension to gravitons of this kind of non-trivial restrictions can be computed with two different methods:
effective field theory operators, or
on-shell helicity amplitude techniques~\cite{1605.07402,1709.04891,2102.11440,2105.11684}.
The second method appears simpler, as
it avoids unnecessarily extending the little-group symmetries of the process into the full Lorentz group.
The decay amplitude of a massive particle $S$ with spin $s$ and momentum $P$
into two massless particles with helicities
$h_1$ and $h_2$ and momenta $p_1$ and $p_2$ is found by writing the momenta $P=p_1+p_2$ as
$p_1^\mu  = \lambda_1 \sigma^\mu \tilde{\lambda}_1$, 
$p_2^\mu = \lambda_2 \sigma^\mu \tilde{\lambda}_2$
in terms of two-component spinors $\lambda^a$,
so that $\sigma^\mu$ are the usual relativistic Pauli matrices.
The amplitude can be written as~\cite{1709.04891}
\beq \label{eq:A}
\mathscr{A}\propto ( \lambda_1^{s+h_2-h_1} \lambda_2^{s+h_1-h_2}) [\tilde\lambda_1^a \epsilon_{ab} \tilde\lambda_2^b]^{s+h_1+h_2},
\eeq
where the term in squared parentheses is the Lorentz scalar built using the anti-symmetric tensor $\epsilon_{ab}$,
and the term in round parentheses indicates the completely symmetric spin $s$ product
of the number of $\lambda_1$ and $\lambda_2$ dictated by their exponents.
Locality demands no negative powers of momenta $\lambda_{1,2}$, so 
helicities must satisfy 
\beq \label{eq:sh1h2}
s\ge |h_1-h_2|.\eeq

\subsection{Decay of a massive particle into two gravitons}\label{sec:S2gg}
Inserting the graviton helicities $h_1,h_2 = \pm 2$ in eq.\eq{A} shows that:
\begin{itemize}
\item[0)] A spin 0 particle can decay into two gravitons, that must have the same helicity  $h_1=h_2$ because of eq.\eq{sh1h2}.
The amplitude of eq.\eq{A} arises from the effective operators\footnote{Simpler
operators do not contribute for the following reasons.
The tadpole operator $\Lambda^4 \sqrt{|\det g|} S$ 
gets cancelled at the minimum of the $S$ potential.
The operator $\Lambda \, \sqrt{|\det g|} \, SR$ does not contribute to $S$ decays into gravitons,
as it also induces a $S$/graviton kinetic mixing:
one needs to diagonalise the action reaching
mass eigenstates via a Weyl field redefinition to  the Einstein frame,
that removes such operator.
Effective operators of the form $\sqrt{|\det g|}  SR^2/\Lambda$ or $\sqrt{|\det g|}  S\, R_{\mu\nu}R^{\mu\nu}/\Lambda$ 
do not contribute since on-shell gravitons in flat space satisfy $R_{\mu\nu}=0$ and thereby $R=0$~\cite{2105.11684,2112.12774}.
The first operator of eq.\eq{SRR} is mediated at one loop by matter fields with coefficient proportional to a $\beta$ function of 4-derivative gravity.
The second operator involving the dual Riemann tensor $\tilde{R}$ can be mediated by one loop of chiral fermions.
}
\beq \label{eq:SRR}
\sqrt{|\det g|}  S\, R_{\mu\nu\rho\sigma}R^{\mu\nu\rho\sigma}/\Lambda,\qquad
\sqrt{|\det g|}  S\, R_{\mu\nu\rho\sigma}\tilde{R}^{\mu\nu\rho\sigma}/\tilde\Lambda\eeq
where $\Lambda, \tilde\Lambda$ are mass scales and $\tilde{R}$ is the dual Riemann tensor.
The first operator contributes to $S$ decay into two gravitons as~\cite{2105.11684}\footnote{Its gravitational wave effects 
have been studied in~\cite{2112.12774} in a parameter range with
$H_{\rm infl}\sim T_{\rm RH}$ that leads to a large $f_{0\rm peak}\sim 10^{23}\,{\rm Hz}$.} 
\beq \label{eq:fund}
\mathscr{A}(S\to h^{++}h^{++}) = \frac{[\tilde\lambda_1\tilde\lambda_2]^4}{2\Lambda\bp^2},\qquad
\Gamma_{\rm GW} =  \frac{M^7}{64\pi \Lambda^2 \bar{M}_{\rm Pl}^4}.
\eeq

\item[1)] The decay amplitude for a spin $s=1$ vector into two gravitons vanishes
because of the anti-symmetry of the last factor in eq.\eq{A}.
The only possible decay of a vector into two identical particles is into spin 1/2 fermions.

\item[2)] A spin 2 particle can decay into two same-helicity gravitons $h_1=h_2$.
The effective operator is a contraction of 
$(\partial_\mu \partial_\nu S_{\alpha\beta})\,{\rm Riemann}^2/\Lambda^3$ and
\beq \mathscr{A} (S\to h^{++}h^{++}) \sim  \frac{[\tilde\lambda_1\tilde\lambda_2]^6 \lambda_1^2\lambda_2^2}{M^2\Lambda^3 \bp^2},\qquad
\Gamma_{\rm GW} \sim \frac{M^{11}}{ 4\pi \Lambda^6 \bar{M}_{\rm Pl}^4 }.
\eeq

\item[3)] A spin 3 particle cannot decay into two gravitons, similarly to a spin 1 particle.

\item[4)] A spin 4 particle can decay into two same-helicity gravitons and (via a lower dimensional operator)
into opposite-helicity gravitons. We do not consider this possibility.
\end{itemize}

\subsection{Decay of a massive particle into a graviton and a SM particle}\label{sec:S'}
A process that produces one graviton $g$ only has a rate suppressed by only two powers of the Planck scale.
Given that we are interested in values of $M$ much above the weak scale,
we can approximate the SM particle as massless and apply eq.\eq{A},
inserting now $h_1=\pm 2$ and $-1\le h_2\le 1$, the possible helicities of a SM particle.
The amplitude shows that:
\begin{itemize}
\item[0)] A massive spin 0 or 1/2 particle cannot decay into a graviton and a SM particle.
\item[1)] A massive spin 1 vector 
can decay as $S \to g \gamma$ where $\gamma$ denotes
an abelian SM vector such that $h_1 = 2$ and $h_2 = 1$.
Effective operators such as
$R^{\alpha\beta\mu\nu}F_{\alpha \beta}S_{\mu\nu}/\Lambda^d$ 
(where the anti-symmetric field $S_{\mu\nu}$ could have mass dimension $d=2$ or $1$)
lead to
\beq\label{eq:Vggamma}
\mathscr{A}(S\to h^{++}\gamma^{+}) \sim \frac{[\tilde\lambda_1\tilde\lambda_2]^4 \lambda_2^2}{M^{3-d}\Lambda^d\bp},\qquad
\Gamma_{\rm GW} \sim  \frac{M^{3+2d}}{ 4\pi \Lambda^{2d} \bar{M}_{\rm Pl}^2}.
\eeq

\item[3/2)] A massive spin 3/2 particle can decay into a graviton and a SM fermion $\psi$ as
\beq\label{eq:Vgravitino}
\mathscr{A}(S\to  h \psi) \sim \frac{[\tilde\lambda_1\tilde\lambda_2]^4 \lambda_2^3}{M^{5/2}\Lambda \bp},\qquad
\Gamma_{\rm GW} \sim  \frac{M^{5}}{ 4\pi \Lambda^{2} \bar{M}_{\rm Pl}^2}.
\eeq

\item[2)] A massive spin 2 vector can decay into $g\gamma$ and into $g h$, 
where $h$ is a massless neutral scalar
(so the decay amplitude into the Higgs is suppressed by the weak scale $v$).
\end{itemize}
In both cases 1) and 2), $S$ can also decay into two SM vectors.
If the decay operator is mediated by a loop of particles with mass $m$ and gauge charge $g$,
one expects $1/\Lambda^2 \sim g/(4\pi m)^2$ and
$\Gamma_{\rm GW}/\Gamma_{\rm SM}\sim (m/g\bp)^2 \lesssim 1$
if gravity is the weakest interaction.

\subsection{Decay of a massive particle into a graviton and a massive particle}\label{sec:2massive}
Finally, we consider a decay $S'\to Sg$ where $S$ and $S'$ are two massive particles with spins $s$ and $s'$~\cite{1709.04891}.
The graviton energy from the decay at rest is $E = (M'^2-M^2)/2M'$.
The decay is allowed if $s+s'\ge 2$ as long as $S$ and $S'$ are either both bosonic or both fermionic.
For low spins this reduces to the spin combinations listed in the previous section.
At larger spins, decays with $|s-s'|> 2$ become allowed, altought suppressed in the non-relativistic limit.

\section{Theories for decay into gravitons}\label{sec:models}
The next key issue is whether in motivated theories the suppression scales
$\Lambda$ of operators for decays into gravitons can have a sub-Planckian value, down to $\Lambda \gtrsim M$
where the range of validity of the effective operator approximation breaks down.
Section~\ref{sec:G} presents theories based on strong gauge dynamics
that achieve  $\Lambda\sim M$ and thereby the maximal gravitational wave abundance.
Section~\ref{sec:BE} presents theories with weak couplings $g\sim 1$, resulting in 
$\Lambda \gtrsim (4\pi)^2 M/g$ a loop factor above $M$.

\begin{figure}[t]
\centering
$$\includegraphics[width=0.95\textwidth]{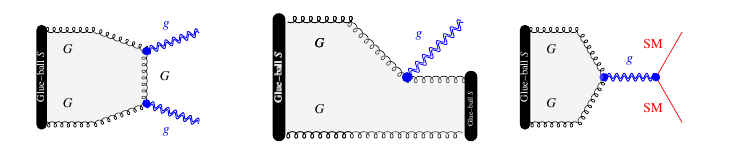}$$
\vspace{-4ex}
\caption{\em\label{fig:DGgravitationalDecays}
Illustrative Feynman diagrams for gravitational glue-ball decays. 
Blue dots denotes gravitational couplings.
SM denotes any Standard Model particle, including gravitons.}
\end{figure}

\subsection{Theories with non-perturbative interactions}\label{sec:G}
Maximal gravitational decays arise in theories with gauge group structure $G_{\rm SM}\otimes G$, 
provided the extra factor $G$ becomes strongly coupled and remains hidden.
Hidden means that no sub-Planckian matter fields couple the two sectors at renormalizable level.
Concretely, this means that no matter field can be charged under both factors $G_{\rm SM}$ and $G$,
and no scalar charged under $G$ can have  a quartic coupling to the SM Higgs.
We will focus our discussion on the simplest possibility: a pure-gauge non-abelian gauge interaction $G$.
This is automatically hidden, as gauge invariance forbids renormalizable interactions with SM particles.
We thereby extend the SM adding an extra pure-glue gauge sector  $G$. The action is 
\beq\label{eq:LDM}
S =  \int d^4x \sqrt{|\det g|} \left[-\frac{\bp^2}{2}R - \frac14 {G}_{\mu\nu}^a G^{\mu\nu a} + \Lag_{\rm SM} + \Lag_{\rm NRO} \right]\eeq
where $G_{\mu\nu}^a = \partial_\mu G_\nu^a- \partial_\nu G_\mu^a- g f^{abc} G^b_\mu G^c_\nu$,
$g$ is the gauge coupling,
and we left implicit the dark $\theta$ term.
The extra gauge vectors interact with SM particles only via gravitational interactions and via possible
non-renormalizable operators $\Lag_{\rm NRO}$ that we assume to be Planck-suppressed,
corresponding to the absence of light matter field charged both under $G$ and under the SM.
Planck-suppressed operators are unavoidably generated by RG running. 
Similarly to QCD, the gauge interaction $G$ confines at some scale $\Lambda$ naturally below the Planck scale.
The confinement phase transition is expected to be of 1st order with slow bubbles~\cite{2105.02840},
so it produces a negligible extra amount of gravitons.
Being a pure glue theory, confinement forms a set of glue-ball composite states with masses $M \sim\Lambda$.
The abundance of glue-balls from pair production of dark massless vectors via gravitational interactions (gravitational freeze-in) followed by confinement is~\cite{2012.12087}:
\begin{itemize}
\item For low masses $M\lesssim\bar{\mpl}(\trh/\bar{\mpl})^{15/4}$, 
 the dark vectors thermalize among themselves before confining into glue-balls with abundance $Y_S\sim(\trh/\bar{\mpl})^{9/4}$,
 neglecting logarithmic corrections from $SSS\to SS$ scatterings.
 We omitted $10^{-2-3}$ factors, that depend on the dimension of the gauge group.
 \item In the intermediate range $\bar{\mpl}(\trh/\bar{\mpl})^{15/4}\lesssim M\lesssim \trh^2/\bar{\mpl}$, 
the glue-balls self-thermalize.
The abundance is similar to previous case.
    \item For larger masses, the dark vectors immediately hadronise and the glue-balls never reach an equilibrium distribution. The resulting abundance is $Y_S\sim N(\trh/\bar{\mpl})^3 e^{-M/\trh}$, 
where $N$ is an enhancement due to the particle multiplicity in $G$-vector jets produced by collisions at energy 
$\trh\gg\Lambda$~\cite{2012.12087}.
\end{itemize}
The lightest glue-ball corresponds to the operator $\Tr\, G^2$: it has spin $s=0$ and we denote it as $S$.
Its decay rates into gravitons and SM particles are estimated to be comparable and have the form of
eq.\eq{GammaGW} for $p=4$:
\via{We expand around the flat-space metric $\eta_{\mu\nu}=\diag(1,-1,-1,-1)$ as $g_{\mu\nu} = \eta_{\mu\nu} + 2 h_{\mu\nu}/\bp$ 
(as in Berends-Gastmans) such that one graviton $h_{\mu\nu}$ couples as
$h_{\mu\nu} T^{\mu\nu} /\bar M_{\rm Pl}$ 
where $T^{\mu\nu} \equiv 2\, {\delta S}/\delta g_{\mu\nu} = T^{\mu\nu}_{\rm SM}+T^{\mu\nu}_{\rm DM} + \cdots$ is the usual energy-momentum tensor.
My code Tensor.nb used the different expansion $g = \eta + \kappa h$ with $\kappa=1/\bp$ and 
$\Lag_{\rm gf}=(\partial_\mu h^{\mu\nu} - c \partial^\nu h /2)^2/2\xi$.
Then the Einstein term gets expanded as
\beq \Lag = k^2 \bigg[ P^{(2)}  -2 P^{(0)}\bigg]_{\mu\nu\rho\sigma}  \frac{h_{\mu\nu}h_{\rho\sigma}}{2}\eeq
so the graviton propagator (up to gauge fixing) is 
\beq D_{\mu\nu\mu'\nu'}= \frac{i}{k^2}\left[ P^{(2)} - \frac{1}{2} P^{(0)}\right]_{\mu\nu\mu'\nu'}  = \frac{i}{2k^2}(\eta_{\mu\mu'} \eta_{\nu\nu'}+\eta_{\mu\nu'}\eta_{\nu\mu'}-
\frac12 \eta_{\mu\nu}\eta_{\mu'\nu'})+ \cdots\eeq
We defined the projectors on the spin 2, spin 1, spin 0 states that sum to unity,
$(P^{(2)}+P^{(1)}+P^{(0)}+P^{(0w)})_{\mu\nu\rho\sigma}=\frac12 (\eta_{\mu\nu}\eta_{\rho\sigma}+\eta_{\mu\sigma}\eta_{\rho\nu})$:
\begin{eqnarray}
P^{(2)}_{\mu\nu\rho\sigma} &=& \frac12 T_{\mu\rho}T_{\nu\sigma} + \frac12 T_{\mu\sigma} T_{\nu\rho}-\frac13 T_{\mu\nu}T_{\rho \sigma}\\
P^{(1)}_{\mu\nu\rho\sigma} &=& \frac12 (T_{\mu\rho}L_{\nu\sigma} +  T_{\mu\sigma} L_{\nu\rho}+
 T_{\nu\rho}L_{\mu\sigma} +  T_{\nu\sigma} L_{\mu\rho})\\
 P^{(0)}_{\mu\nu\rho\sigma} &=& \frac13 T_{\mu\nu}T_{\rho \sigma}\\
 P^{(0w)}_{\mu\nu\rho\sigma} &=& L_{\mu\nu}L_{\rho \sigma}
\end{eqnarray}
where
\beq T_{\mu\nu} = \eta_{\mu\nu} - k_\mu k_\nu/k^2, \qquad L_{\mu\nu}=k_\mu k_\nu/k^2.\eeq
} 
\beq \label{eq:estimate}
\Gamma_{\rm SM}=
\Gamma(S \to {\rm SM}\,{\rm SM}) \sim g_{\rm SM} \frac{M^5}{4\pi \bp^4} ,\qquad 
\Gamma_{\rm GW}=
\Gamma(S\to gg) \sim \frac{M^5}{4\pi \bp^4}
\eeq
where $g_{\rm SM}$ is number of SM degrees of freedom. 
A precise prediction of the decay rates is prevented by
uncertainties due to the non-perturbative nature of the glue-ball state,
and by Planck-suppressed operators in $\Lag_{\rm NRO}$, that provide contributions comparable to Einstein gravity.
The estimates in eq.\eq{estimate} are justified as follows.

An Effective Field Theory approach restricts the coupling of the lightest glue-ball $S$
to  {\em soft}  gravitons.
The dominant operator would be the tadpole
$\sqrt{|\det g|} S$, with coefficient computable from the covariantization of the 
matrix element $\bAk{0}{\Tr G^2}{S}$, known from lattice computations~\cite{hep-lat/0510074}.
However, as discussed in section~\ref{sec:S2gg}, the tadpole vanishes at the minimum and does not lead to a gravitational decay.
Going to higher orders, the dominant effect is given by eq.\eq{fund}
with expected suppression scale $\Lambda \sim M$, leading to the rates in eq.\eq{estimate}. 

However, higher order operators are relevant and the EFT expansion breaks down, as $S$ decays into
{\em hard} gravitons with energy $E = M/2$.
Their production probes the glue-ball inner structure  and thereby depends on unknown non-perturbative form factors.\via{One would need to compute the matrix element.
Simpler ones are known from lattice computations, such as 
$f_{0S}\equiv \langle 0 | {\rm Tr}\,{\cal G}_{\mu\nu} {\cal G}^{\mu\nu}| 0^{++}\rangle$
computed as
$4\pi \adc f_{0S}\approx 3 \mDG ^3$.
However, we need more complicated ones, $\bAk{h_{\mu\nu} (k) h_{\mu'\nu'}(k')}{H_{\rm int}}{S}$.
Diagrams with $s$-channel exchange of one graviton (that next couples to other two gravitons or two SM particles)
contribute to the amplitude as $\mathscr{A}= -i {\cal T}/\bar{M}_{\rm Pl}^2 s$  where ${\cal T} = T_{\mu\nu} T^{\mu\nu}$.
So one needs 
\beq    \bAk{0}{T_{\mu\nu}}{S(p)} =  f_1(p^2) \eta_{\mu\nu} + f_2(p^2) p_\mu p_\nu \eeq
which is partially known~\cite{0808.3151,1707.05380}.
The other $T_{\mu\nu}$ is trace-less, up to one loop effects, due to scale invariance.
The effective operator $S$ would give $f_1=1$, $f_2=0$
In the presence of $f_2\neq 0$ one presumably gets the unsuppressed estimate of eq.\eq{estimate}.
In other diagrams the two gravitons directly couple to the glue-ball, so one needs matrix elements of the form
\beq  \bAk{0}{\int d^4x \, {\rm T}\,  T_{\mu\nu}(x/2) T_{\mu'\nu'}(-x/2) e^{iq\cdot x}}{S}\qquad
q=(k-k')/2\eeq
which is unknown.
Only the total amplitude is gauge invariant.}
Going beyond the EFT limit, the $S$ decay rate is estimated as follows.
If a non-relativistic description of $S$ as a bound state with wave function $\psi_S$ were applicable,
the glue-ball gravitational decay would be computable as a gravitational collision of dark gluons,
as illustrated by the Feynman diagrams in fig.\fig{DGgravitationalDecays}.
Given that non-perturbative couplings lead to relativistic constituents, 
we approximate the collision speed as the speed of light, obtaining
\beq \Gamma(S\to gg, {\rm SM}\,{\rm SM}) \approx |\psi_S|^2  \sigma(GG\to gg, {\rm SM}\,{\rm SM}) \approx \frac{M^5}{4\pi \bp^4}.\eeq
Indeed the squared wave-function $\psi_S$ of the bound state is roughly given by its inverse volume,
$|\psi_S|^2 \sim M^3$,
and the gravitational cross-section of two dark gluons $G$ into two gravitons or two SM particles is
 $\sigma(GG \leftrightarrow gg,{\rm SM}\,{\rm SM}) \sim s/4\pi \bp^4$.
 Graviton-mediated cross sections into SM particles are computed e.g.\ in~\cite{2012.12087}. 
No extra suppression is expected, as
the decay of a massive spin 2 particle into two gravitons is allowed by symmetries as discussed in section~\ref{sec:S2gg}.
Non-renormalizable operators, such as $|H|^2 G_{\mu\nu}^a G^{\mu\nu a}/\bp^2$, can contribute at comparable level.
In conclusion, the expected amount of gravitational waves from glue-ball decays is shown by the middle panel of fig.\fig{GW}.
The dot-dashed green curve shows the expected glue-ball  abundance.

\medskip

\medskip

Each group $G$ predicts a specific glue-ball spectrum that contains extra heavier glue-balls. 
Those with mass $\ge 2M$ can decay via gauge interactions into $SS$, negligibly contributing to gravitational waves.
Some models and computations predict at least one exception:
a resonance $S'$ with spin $s'=2$ and mass $M' \approx 1.5 M$~\cite{2206.14826}. 
Its lightness would imply that $S'\to SS$ is kinematically blocked, so that
these two lighter glue-balls are co-stable.
In such a case $S'$ can decay emitting one graviton $g$ as $S' \to S g$
(allowed according to section~\ref{sec:2massive}), with decay rate estimated as
\beq \Gamma'_{\rm GW}=\Gamma(S'\to S g) \sim \frac{M'^3}{4\pi \bp^2} \left(1-\frac{M^2}{M'^2}\right).\eeq
It is suppressed by only $p=2$ powers of $\bp$ because one graviton only is emitted
(a $S'$ with spin 1 would instead decay as $S'\to S gg$).
This kind of earlier $S'$ decays produce extra gravitational waves at lower frequencies.
The resulting spectrum gravitational waves can feature multiple peaks,
observable in part of the parameter space.
An example is shown in fig.\fig{GW2}: the faster $S'\to S g$ decay with $p=2$
gives gravitational waves peaked around $\sim 10^{13}\,{\rm Hz}$. 
These get mildly suppressed by the later $S\to gg,{\rm SM}\,{\rm SM}$ decay with $p=4$
that gives GW peaked around $10^{15}\,{\rm Hz}$, about $\bp/M$ higher.

\medskip

Some pure gauge theories predict extra special very long lived glue-balls $S''$,
proposed in~\cite{2012.12087} as gravitational dark matter candidates.
For example, the group $G=\SU(N)$ predicts C-odd $GGG$ glue-balls with slow gravitational decay rate $\Gamma_{S''} \sim M^9/\bp^8$
arising from Planck-suppressed operators.
The group $G =\SO(N)$ with even $N$ predicts glue-balls odd under an $\SO(N)$ parity
with decay rate $\Gamma_{S''} \sim M (M/\bp)^{2N-4}$.
Such longer-lived glue-balls match the dark matter density if their abundance is $Y_{S''} \sim \eV/M''$, much below
the $Y_{S,S'} $ that lead to gravitational waves at observable level (their abundance was estimated in~\cite{2012.12087}. See also~\cite{2307.06586} for an alternative estimate).
Naive hadronization models suggest $Y_{S''} /Y_{S,S'} \sim 0.76^N$~\cite{2012.12087}. 
Some other groups don't predict longer-lived glue-balls.


\subsection{Theories with perturbative interactions}\label{sec:BE}
The previous discussion suggests that observing gravitational waves around optical frequencies
might allow to discover `gravitational particles', with mass a few orders of magnitude below the Planck scale, 
that slowly decayed with a non-negligible branching ratio into gravitons. 
We conclude discussing which theories 
contain plausible candidates for such states.

\medskip

\begin{figure}[t]
\centering
$$\includegraphics[width=0.95\textwidth]{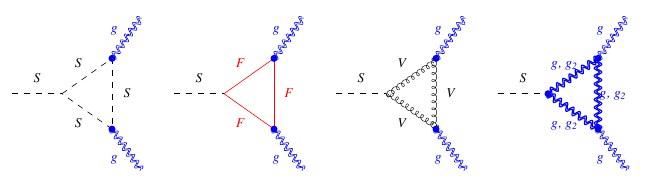}$$
\vspace{-1ex}
\caption{\em\label{fig:DGgravitationalDecaysLoop}
Feynman diagrams for gravitational decays of a Higgs-like particle $S$.}
\end{figure}

We start considering the pessimistic case where operators such as eq.\eq{SRR} arise at one loop level,
from the first three Feynman diagrams in fig.\fig{DGgravitationalDecaysLoop},
by exchanging a scalar (possibly the same scalar $S$) or an extra fermion $F$ or vector $V$.
The scale $\Lambda$ that suppresses the resulting effective operator 
$S\, R_{\mu\nu\rho\sigma}R^{\mu\nu\rho\sigma}/\Lambda$
is estimated via dimensional analysis as 
\beq \Lambda \sim (4\pi)^2 m/g,\label{eq:Lambdaesimate}\eeq
where $m$ is the particle in the loop and $g$ its dimension-less coupling to $S$
(either $g m S^3$, $g S\bar F F$ or $g m S V_\mu^2$).
A precise expression for the coefficient of the scalar operator
is obtained in the limit $m\gg M$ where the operator approximation holds
recalling that the coefficient $\gamma$ of the Riemann squared operator 
\beq \label{eq:Riemann2}
\sqrt{|\det g|} \gamma R_{\mu\nu\rho\sigma}R^{\mu\nu\rho\sigma}\eeq
at one loop level runs as~\cite{hep-th/9510140,1308.3398,2112.12774}
\beq
\frac{\partial\gamma}{\partial\ln \mu} =
 \beta_\gamma^{\rm gravity}+\beta_\gamma^{\rm matter}\qquad\hbox{with}\qquad
\beta_\gamma^{\rm matter}=
 \frac{52 N_V - 7N_F -4N_S}{720 (4\pi)^2}\eeq
in the presence of  $N_S$ real scalars, $N_F$ Weyl fermions, $N_V$ vectors, generically denoted as `matter'.
Indeed, the operator of eq.\eq{SRR} arises when $S$ couples as a Higgs to these matter particles,
giving them  $S$-dependent masses that act as an IR cut-off in the RG running of $\gamma$.
The maximal operator coefficient arises when the masses of these particle arise solely 
from the vacuum expectation value of $S$ as $m_F = y \med{S}$, $m_S = \sqrt{\lambda}\med{S}$, $m_V = g \med{S}$,
where $y$ generically denotes Yukawa couplings, $g$ denotes gauge couplings and $\lambda$ quartic couplings.
In this case the operator arises with coefficient~\cite{2112.12774}
\beq  \label{eq:loopO}
 \frac{1}{\Lambda}=\frac{\beta_\gamma^{\rm mat}}{\med{S}}
\eeq
in agreement with the estimate of eq.\eq{Lambdaesimate}.
Notice that the cubic scalar coupling in the first diagram of fig.\fig{DGgravitationalDecaysLoop}
is induced by the quartic coupling as $\lambda\med{S}$,
and that the $m\gg M$ operator approximation is not applicable if the scalar in the loop has the same mass $M$
as the external scalar $S$.

The $S\, R_{\mu\nu\rho\sigma}\tilde{R}^{\mu\nu\rho\sigma}/\tilde{\Lambda}$ operator
arises from one loop exchange of fermions with Yukawa couplings to a scalar $S$
that affects the phase of the fermion mass.
The operator coefficient can be extracted from the gravitational chiral anomaly, as
a rephasing $F_i \to e^{i c_i S/\med{S}}F_i$ of Weyl spinors $F_i$
can remove the coupling to $S$, while generating
the operator with coefficient~\cite{Fujikawa:1980eg,2112.12774} 
\beq\label{eq:Lambdat}
 \frac{1}{\tilde{\Lambda}}= \frac{\sum_i c_i}{24(4\pi)^2\med{S}}.\eeq
 Eq.s\eq{loopO} and\eq{Lambdat} mean that  the pure gauge theory of section~\ref{sec:G} where
$g \sim (4\pi)^2$ and $m_V \sim M$ provides the maximal $S$ decay rate into gravitons.
If instead $1/\Lambda$ is loop suppressed, the decay width into gravitons $\Gamma_{\rm GW}$
gets suppressed by $ 1/\Lambda^2 \sim g^2/(4\pi )^4m^2$.
The amount of gravitational waves can remain large, despite that $\Gamma_{\rm GW}$ is
now loop suppressed, provided that other decay channels such as $\Gamma_{\rm SM}$ are suppressed down to a similar level.
A first danger is that the $S$ couplings to matter can also allow for $S$ decays into pairs of matter particles at tree level,
that would suppress the $S$ branching ratio into gravitons.
Such decays are kinematically blocked if $m_{S,F,V} \ge M/2$.
The condition is automatically satisfied in the minimal theory where $S$ itself is the only matter particle in the loop.
In such a case, the $S\to gg$ decay arises provided that the $S$ potential has a cubic $S^3$ term when expanded around the minimum.
The second danger is decays into SM particles, that are much lighter than $M$.
Then, the possibility of a small $\Gamma_{\rm SM}$ 
depends on whether the $S,F,V$ matter in the loop directly interacts with SM particles.
 

\medskip

Let us consider {\em grand unified theories}, 
where a unification gauge group breaks to the SM gauge group
via the Higgs mechanism  at mildly sub-Planckian energies $\gtrsim 10^{15}\GeV$, as demanded by proton decay bounds.
Candidates for gravitationally-decaying scalars $S$ are the SM singlets in the 24 of SU(5), in the 16 of SO(10), in the 27 of $E_6$.
Indeed unification theories predict, in addition to SM particles,
extra heavy vectors and scalars (and possibly fermions) that receive masses from $S$
and that thereby mediate at one loop level the operator in eq.\eq{SRR}.
These heavy GUT particles have gauge-like couplings $g_{\rm GUT}\sim 1$ to $S$,
but also to the SM particles.
SM particles are not directly coupled to $S$, because they do not acquire GUT-scale masses.
If $m_{S,F,V}  \ge M$ the decays of $S$ into SM particles
are partially suppressed, by a 4-body phase space or by a 1-loop factor.
The resulting partially small $\Gamma_{\rm SM} \sim g^6 M /(4\pi)^5$ leads to an
amount of gravitational waves that falls between the CMB/BBN bound $\Omega_{\rm GW}\sim 10^{-6}$
and the solar background $\Omega_{\rm GW}\sim 10^{-15}$, depending on $M$.
An Higgs-like scalar lighter than the particles to which it gives mass can arise e.g.\
if the unification symmetry is broken dynamically via the Coleman-Weinberg mechanism.

\subsection{Extensions of Einstein gravity}\label{sec:EE}
Better candidates for gravitationally-decaying particles can be found in extensions of Einstein gravity.
We here present a quick overview of some possibilities, without performing detailed computations.

Theories with (discrete or continuous)
{\em extra dimensions} predict a (finite or infinite) number of massive Kaluza-Klein excitations of gravitons $g_n$ 
as well as extra particles. These include extra scalars: the extra dimensional `radion' 
components of the graviton $g_{MN}$, and extra `moduli'.\footnote{A different
phenomenon, production of gravitational waves from stochastic inhomogeneities in theories with extra dimensions, 
was considered in~\cite{1704.07392,1906.11652}.}
The zero modes of such fields control the size and shape of the extra dimensions.
Thereby such fields act as geometric Higgs bosons, affecting the masses of matter fields.
Eq.\eq{loopO} means that the $S\to gg$ operator of eq.\eq{SRR} arises at the compactification scale.

The lower Kaluza-Klein excitations $g_n$ of the graviton $g_0$
have spin 2, so they could decay into gravitons such as
\beq g_1 \to g_0 g_0,\qquad g_2 \to g_1 g_0.\eeq
This kind of decays can happen if two conditions are satisfied.
First, the extra dimension must be not translationally invariant, to avoid conservation of KK number.
Second, the higher dimensional  gravitational action cannot have
the minimal Einstein form (curvature plus vacuum energy), and must 
contain 4-derivative Riemann squared terms (again unavoidably generated by one loop corrections).
These non-minimal gravitational terms are necessary because  the effective operators identified in section~\ref{sec:Amp} 
all involve 4 derivatives.\footnote{Furthermore, if the gravitational action only contains 
minimal Einstein terms and vacuum energy terms (possibly including background contributions localised along sub-spaces such as branes and orbifold fixed points), graviton KK modes are described by eigenfunctions of the Laplacian, 
and different modes are orthogonal when weighted by the background metric.
The cubic couplings among different KK graviton modes $n_{1,2,3}$ are proportional to wave-function overlaps $I_{n_1 n_2 n_3}$.
We live in a special compactification, with energies tuned
such that the 4-dimensional vacuum energy nearly vanishes
(this might require a warping or negative-energy non-dynamical objects).
Then  the graviton 0 mode $g_0$ is massless and its wave-function is constant, when weighted by the metric~\cite{2306.05456}.
This implies the vanishing of some
 overlap integrals, such as $I_{100}\propto I_{10}=0$ and $I_{210}\propto I_{21}=0$.
}



Warping is one plausible source of breaking of extra-dimensional translational invariance.
However, according to the AdS/CFT duality, 
Kaluza Klein graviton excitations in a warped geometry can be re-interpreted as spin 2 glue-balls of some conformal field theory
in 4 dimensions.
More general holographic techniques can similarly approximate the spectrum of scalar and vector glue-balls~\cite{2206.14826}
arising from 4-dimensional strong dynamics, similar to the theories of section~\ref{sec:G}.


\smallskip

A related possibility is that the SM fields are confined around some value of the extra-dimensional coordinates,
thereby approximatively living along a 3+1 dimensional brane in extra dimensions.
In such a case, the 4-dimensional effective field theory also contains a set of massive `branon' scalars,
describing fluctuations in the position of the brane in the extra dimensions.
The branon potential energy can contain a self cubic term, leading to decays into
gravitons at one loop level from the first diagram in fig.\fig{DGgravitationalDecaysLoop}.

\medskip

Given that decays into gravitons need Riemann squared operators,
an interesting possibility is the most general action in 3+1 dimensions containing
 terms with positive mass dimension only,
as it gives a renormalizable theory of quantum gravity~\cite{Stelle:1976gc}.
This is known as {\em 4-derivative gravity} because
$R^2$ and $R_{\mu\nu}^2$ terms in the action imply a graviton kinetic term with 4 derivatives.
As a result the theory
contains the usual graviton, a possibly problematic spin 2 ghost, and
the spin 0 gravi-scalar also present in Starobinsky $R^2$ inflation.
Such extra states could decay into gravitons.
However these decay rates vanish at tree level~\cite{1804.04980},
because the Gauss-Bonnet term is topological in 3+1 dimensions.
A non-vanishing rate is expected at loop level.\footnote{4-derivative theories predict a different big source of gravitational waves:
UV divergences are made renormalizable at the price of enhancing IR divergences,
such that energies above the ghost mass are released via graviton bremsstrahlung~\cite{1808.07883}.}

\smallskip

More in general, the Planck mass might be dynamically induced by the vacuum expectation value of a scalar $S$, dubbed {\em Planckion}.
As it acts as the Higgs for gravity, one expects $S\to gg$ similar to eq.\eq{loopO},
mediated at loop level by a purely gravitational loop (4th diagram in fig.\fig{DGgravitationalDecaysLoop}).
This loop is computable in renormalizable 4-derivative gravity that predicts
$\beta_\gamma^{\rm gravity}=-413/180(4\pi)^2$~\cite{hep-th/9510140,1308.3398}.

\medskip

If {\em supersymmetry} exists and is broken around the Planck scale, 
the minimal $N=1$ gravitational super-multiplet 
contains the massless graviton,
a massive spin 3/2 gravitino and two scalars
(remnants of the chiral super-multiplet `eaten' by the massless gravitational super-multiplet),
such that the number of bosonic and fermionic degrees of freedom match.
The gravitino does not decay into a graviton and a SM fermion.
Theories where supersymmetry is broken in an hidden sector
are often considered to avoid problematic spectra of super-particles.
In this context, the two extra scalars can be gravitationally decaying particles.

\medskip

{\em Strings} $X_M(\tau,\sigma)$ in 10 dimensions have been proposed as theories of quantum gravity~\cite{Iba}.
A huge number of possible compactifications 
can lead to the gravitationally-decaying particles typical of extra dimensions,
but hugely reduce the string predictivity.
For example one of the many possibilities is that the theories of section~\ref{sec:G}  with a maximal gravitational wave abundance
could arise with hidden group $G=E_8$ from the heterotic string~\cite{Iba}.
To avoid getting lost in a plethora of possibilities, lets us focus on the most characteristic fields
arising in the QFT limit of strings: fields with the quantum numbers of $X_M X_N$, 
that decompose into the graviton $g_{MN}$ (symmetric two-index gauge field) accompanied by
a scalar and by the Kalb-Ramond field $B_{MN}$ (anti-symmetric two-index gauge field).
A variety of sources can make the KR field massive~\cite{Iba,2309.02485}.
Then, its 4-dimensional component is dual to a pseudo-vector $S_{\mu\nu}$,
with suppressed couplings to matter that include axion-like couplings, magnetic-like couplings to fermions,
and interactions with vectors and gravitons.
The KR field could decay into a graviton and a SM vector as in eq.\eq{Vggamma}.
The higher-dimensional components of $B_{MN}$ are dual to pseudo-scalars, that can decay in two gravitons as in eq.\eq{fund}.
Other stringy candidates for gravitationally-decaying scalars are moduli.
In particular, chiral fermions and CP violation in 3+1 dimension can arise in string theories from compactification
on extra dimension with complex structure. Toroidal-like compactifications can lead to a QFT with 
a stringy ${\rm SL}(2,\mathbb{Z})$ symmetry~\cite{Iba}. 
Its modulus $\tau$ is equivalent to a combination of scalars that spontaneously break CP.
Thereby $\tau$ affects the phase of fermion masses, and is
expected to acquire a gravitational decay width as in eq.\eq{Lambdat}, plus a Planck-suppressed
width into SM particles.

\medskip

Trying to be more general than strings, graviton amplitudes with good properties seem to
need heavy states along one or more {\em Regge trajectories}, see e.g.~\cite{2312.07652,2406.12959}.
This kind of states could decay gravitationally, be sub-Planckian and maybe generalise the string excitations.

\smallskip

Alternatively, these attempts could just be an approximation to quantum gravity analogous to how Regge trajectories approximate QCD.
The analogy suggests that a deeper quantum gravity theory might feature gravitons and
space-time emerging out of (unknown) fermionic dynamics.
Such an hypothetical  quantum gravity theory could feature, around the Planck scale, massive 
spin 3/2 `excitations' of SM fermions, that could decay into a graviton and a SM fermion
as discussed in section~\ref{sec:S'}.

\smallskip

Finally, a signal could arise from hypothetical 
primordial black holes with mass $M$ mildly above the Planck scale.
These would have evaporated with rate $ g_{\rm SM} \pi \bp^4/80 M^3$ 
and branching ratio ${\rm BR}_{\rm GW}\approx 1/g_{\rm SM}\sim 0.01$ into gravitons,
with a different spectrum~\cite{2302.10188}.

\smallskip

In conclusion, this section briefly outlined a range of possibilities without performing detailed studies, as its aim is to demonstrate that plausible candidates for mildly sub-Planckian states with significant gravitational decays 
seem easily found in plausible theories.
The intent is not to focus on a particular theory or string compactification, 
but rather to highlight that a new observational window --- optical gravitational waves --- 
could provide some experimental insight into a broad class of intriguing theories that might otherwise remain beyond the reach of empirical testing.

\section{Conclusions}\label{concl}
We proposed theories where relic gravitational waves around optical frequencies can have densities
mildly below the CMB/BBN bound on extra radiation,
$\Omega_{\rm GW}\lesssim 10^{-6}$, despite that gravitons interact less than photons.
This can happen if some long-lived particle $S$ has a decay channel {\em opened} by gravitons,
and thereby proceeding with a Planck-suppressed rate.
The cosmology is computed in section~\ref{sec:cosmo}, finding that a particle that decays at a temperature $T_{\rm peak}$
much lower than its mass $M$ gives a specific graviton spectrum, today peaked around optical
frequencies $\sim T_0 M/T_{\rm peak}$.
Simple analytic approximations are provided.
The detailed shape of the gravitational wave peak depends on whether $S$ decays as a sub-dominant component,
or while dominating the cosmological abundance.
In the latter case, a decay with branching ratio into gravitons given by the inverse of the number of SM degrees of freedom
(as naively expected in various theories) 
implies a gravitational wave abundance just below current CMB/BBN bounds.
\begin{itemize}
\item Section~\ref{sec:S2gg} shows that a particle with spin $s=0$ or $2$ can decay into two gravitons,
with rate suppressed by $p\ge 4$ powers of the Planck mass. 
The middle panel of fig.\fig{GW} shows that gravitational wave signals can be observable
even if the $S$ abundance is low and/or $M$ is mildly sub-Planckian.

\item Section~\ref{sec:S'} shows that a particle with spin $s=1$ or $2$ can decay into a graviton and a SM particle
with rate suppressed by $p\ge 2$ powers of the Planck mass.
The left panel of fig.\fig{GW} shows that gravitational wave signals can be observable
provided that the $S$ abundance is mildly small.

\end{itemize}
Section~\ref{sec:models} presents plausible theories where such decays generate gravitational waves with observable abundance.
The maximal abundance arises in theories with a hidden extra gauge interaction that becomes non-perturbative
and confines at some scale $\Lambda$.
A simple case is an extra pure gauge sector 
(thereby automatically hidden) that forms a spectrum of glue-ball states with mass $M\sim\Lambda$.
Such states decay gravitationally into a characteristic gravitational wave spectrum with multiple peaks,
as exemplified in fig.\fig{GW2}.

Moving from theories at strong coupling to 
perturbative SM-like couplings $g\sim 1$,
the decay rate into gravitons gets loop suppressed.
The amount of gravitational waves remains detectable provided that the decay rate into SM particles too is suppressed.
Plausible candidates for gravitationally decaying particles are found, under favourable but plausible conditions, 
in a variety of plausible theories mildly below the Planck scale.
As discussed in sections~\ref{sec:BE}, \ref{sec:EE}, these include gauge unification, extra dimensions, supersymmetry, strings.

\smallskip

These theories were considered before the Large Hadron Collider, that however found no new physics around the weak scale.
Thereby they remain in a limbo: these new physics plausibly lies around Planckian energies, 
many orders of magnitude above the energies where physics is an experimental discipline.
Optical gravitational waves might offer a new window and maybe even some spectroscopical information
on gravitationally decaying particles mildly below the Planck scale. 
This possibility appears as interesting as colliders.
Detecting optical gravitational waves  would need experiments
able of reaching sensitivity to $\Omega_{\rm GW}\sim 10^{-6}$, and improving possibly
down to the ultimate solar astrophysical background, $\Omega_{\rm GW}\sim 10^{-15}$~\cite{gr-qc/0406089,2407.18297}.
Longo lo cammino ma grande la meta~\cite{Branca}.


\small\footnotesize

\paragraph{Acknowledgements}
We thank Gian Giudice, Andrew Long, Michele Redi, Alberto Salvio, Riccardo Torre, Michael Zantedeschi 
for discussions,
and Anish Ghoshal who collaborated in an early phase.
We do not thank Artificial Intelligences, because all most advanced systems
gave the wrong answer to:
``can a spin 0 particle decay into two gravitons?''.
A.S.\ thanks Duccio Pappadopulo for 
dedicated prompts (that allow to get also the correct answer, with roughy $50\%$ probability)
and Riccardo Rattazzi, who correctly answered to the prompt.
 G.L.\ is supported by the Generalitat Valenciana APOSTD/2023 grant CIAPOS/2022/193.

\end{document}